\documentclass[runningheads]{llncs}

\newcounter{noteDMctr} 

\newif\ifEditMode

\EditModetrue

\usepackage{aliases}
\usepackage{preamble}

\begin{document}
\sloppy

\title{Cross-Rollup MEV: \\Non-Atomic Arbitrage Across L2 Blockchains\thanks{Some of this work was performed while the authors were at Matter Labs.}}

\titlerunning{Cross-Rollup MEV} 

\author{Krzysztof Gogol\inst{1} \and
Johnnatan Messias\inst{2} \and
Deborah Miori \inst{3} \and
Claudio Tessone  \inst{1,5}  \and
Benjamin Livshits \inst{4}}

\authorrunning{Gogol et al.}

\institute{University of Zurich \and
Unaffiliated \and
University of Oxford \and
Imperial College London \and
UZH Blockchain Center}

\maketitle
\begin{abstract}

This study quantifies the potential non-atomic MEV on Layer-2 (L2) blockchains by measuring the arbitrage opportunities between cross-rollup and DEX-CEX. Over recent years, we observe a shift in trading activities from Ethereum to rollups, with swaps on rollups occurring 2--3 times more frequently, albeit with lower trade volumes. By analyzing the costs of swap on L2s and price discrepancies cross-rollup and DEX-CEX, we identify more than~\num{500000} unexplored arbitrage opportunities. In particular, we find that these opportunities persist, on average, for~10 to~20 blocks, necessitating the modification of the Loss Versus Rebalancing (LVR) metric to prevent double-counting. Our findings indicate that the arbitrage opportunities in Arbitrum, Base, and Optimism range between~0.03\% and~0.05\% of the trading volume, while in the ZKsync Era it fluctuates around~0.25\%.
 
\end{abstract}


\section{Introduction}
    \label{sec:intro}

\begin{figure}[t]
\centering
\includegraphics[width=0.7\textwidth]{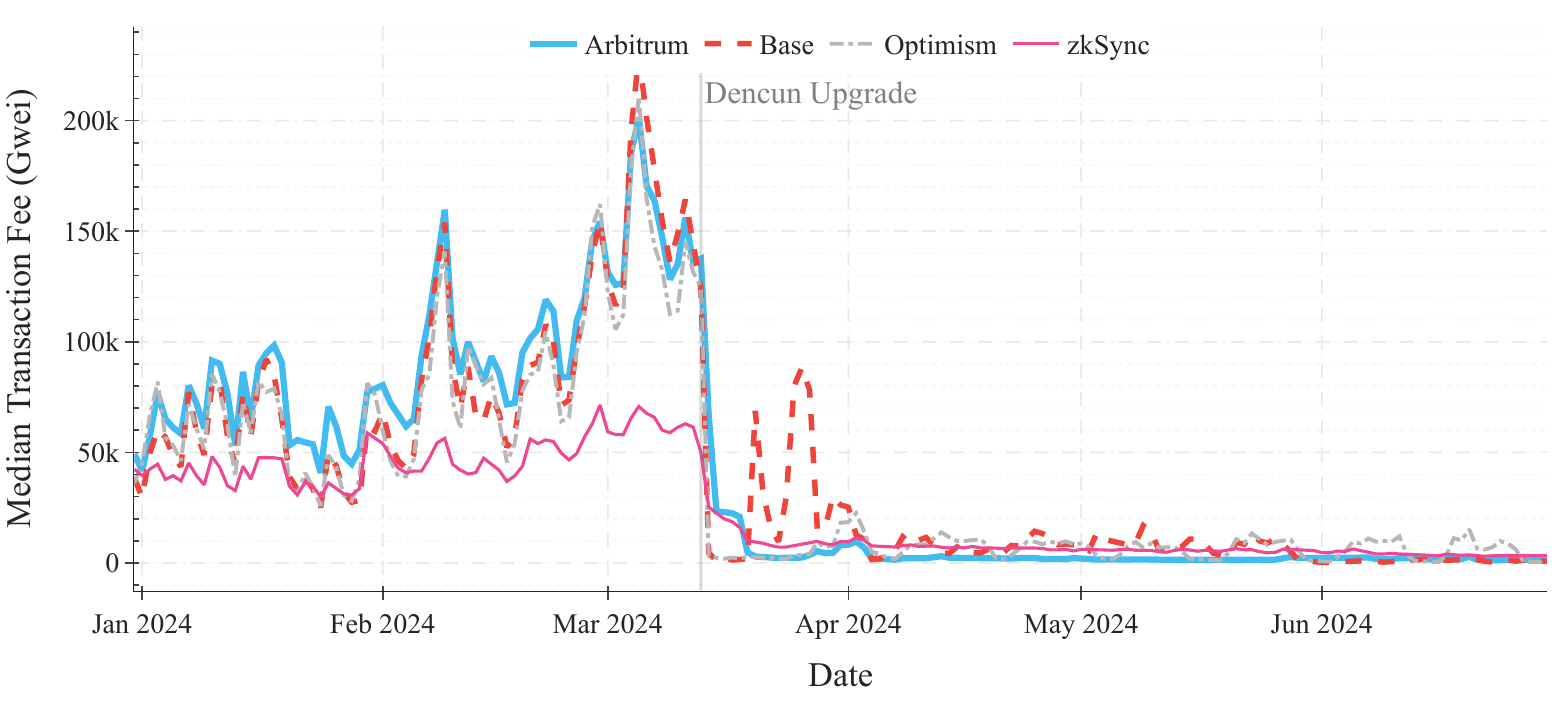} 
\caption{Median transaction fees, measured in Gwei, for selected Ethereum rollups, pre- and post-Dencun Upgrade on March~13\tsup{th},~2024.}
\label{fig:rollup-fees}
\end{figure}

Decentralized Finance (DeFi) is transitioning towards rollups, the Layer-2 (L2) scaling solutions for Ethereum. The primary reason is that rollups provide equivalent security guarantees as Ethereum while significantly reducing gas fees~\cite{Thibault2022SoKRollup}. Rollups offload complex computations outside the Ethereum network, storing only compressed batches of transactions on-chain. In March~2023, Ethereum underwent the Dencun upgrade~\cite{2024EthereumRoadmap}, the most substantial change since the Merge, which further lowered gas fees on rollups, as shown in Figure~\ref{fig:rollup-fees}. The Dencun upgrade introduced Proto-Danksharding and blobs~---~temporary data storage specifically designed to optimize rollups performance~\cite{Eth-Danksharding}.

The trading volume on Decentralized Exchanges (DEXs) within L2s is increasing, with certain rollups surpassing Ethereum's daily swap counts~\cite{2024DeFiCategories,2024L2BeatTVL}. Then, DEXs within L2s present new opportunities for arbitrageurs aiming to capitalize on price disparities across rollups or between rollups and Centralized Exchange (CEX). However, while arbitrage on Ethereum is currently facilitated through Maximal Extractable Value (MEV) boost auctions~\cite{heimbach2024nonatomic}, such mechanisms are yet unavailable in rollups. Most L2s rely on centralized sequencers to produce blocks and aggregate them into Layer-1 batches, employing the first-come-first-served rule. In addition, block production on L2s usually takes less than 2 seconds, as opposed to Ethereum's~12-second block time~\cite{Thibault2022SoKRollup}. The combination of reduced gas costs and faster block production allows arbitrageurs to exploit minor price differences at a faster pace.

Arbitrage can be classified into two types: i) atomic arbitrage within DEXs on the same blockchain (cycling arbitrage~\cite{torres2024rolling}) or ii) non-atomic arbitrage involving CEX or DEX on a different blockchain~\cite{heimbach2024nonatomic}. Currently,~71\% of the MEV transaction volume on Ethereum is related to arbitrage, and in excess of~60\% of that arbitrage volume comes from nonatomic transactions~\cite{dashboard@EighenPhi,dashboard@Flashbots,heimbach2024nonatomic}.
Cycling arbitrage on L2 can be executed as atomic transactions~\cite{torres2024rolling}, but non-atomic arbitrage faces a risk of state (price) drift. With the significantly reduced gas fees following the Dencun upgrade (see Figure~\ref{fig:rollup-fees}), non-atomic arbitrage can be exploited by posting a large amount of prospective arbitrage transactions that are reverted if unsuccessful. The number of revered transactions on L2s increased from~2\% to~6\% after the Dencun upgrade~\cite{dashboard@reverted}. Another strategy used by L2 arbitrageurs is the physical location of centralized sequencers, a service sometimes provided by DeFi solvers~\cite{chitra2024analysisintentbasedmarkets}. Transactions originating from the same data center as the centralized sequencer are likely to be executed first due to shorter network delays~\cite{Daian@Flashboys2,lewis2014flash,wah2016latency}. However, risks associated with centralized sequencers related to censorship remain a concern within L2 DeFi. Additional risks include the transaction finality, the time period until an L2 transaction becomes irrevocable from the rollup (soft finality) or the underlying L1 blockchain (hard finality)~\cite{yee2022shades}.

Rollups can be categorized into Optimistic and Zero-Knowledge~(ZK) rollups based on transaction validity and proof mechanisms~\cite{gangwal2022survey}. Despite the seven-day hard finality period, optimistic rollups have gained user adoption more rapidly due to better Ethereum Virtual Machine~(EVM) compatibility. However, ZK-rollups offer higher transaction compression, throughput, and faster hard finality, ranging from minutes to hours. The aforementioned factors~—~reduced gas fees, absence of MEV auctions, varying finality times, and faster block production~—~contribute to distinct arbitrage dynamics on rollups, which requires further investigation.

With the emergence of decentralized sequencers~\cite{motepalli2023sok}, the adverse practices associated with front-running MEV on Ethereum~\cite{Daian@Flashboys2,qin2021quantifying,Torres-Frontrunning@SP} transfer to L2 DeFi. To address this concern, a time-boost mechanism within sequencers has been proposed~\cite{mamageishvili2023buying}. This mechanism is specifically designed to allow only back-running MEVs while preventing front-running. 
Front-running MEV can enable various forms of adversarial actions, such as sandwich attacks on traders~\cite{Heimbach_2022,messias2023dissecting} and Just-In-Time (JIT) liquidity attacks on Liquidity Providers (LPs)~\cite{zhou2021justintime}. In contrast, back-running MEV enables arbitrage in DEXs and liquidations at lending protocols. Faster arbitrage execution benefits traders, as it eliminates price discrepancies between trading venues. Additionally, MEV time-boost auctions can generate new revenue for sequencer operators. 


\subsection{Contributions}
This work contributes to the research on DeFi and MEV on L2s by quantifying the potential of non-atomic arbitrage on rollups. We conduct an empirical investigation into price disparities cross-rollups and measure the decay time. Our findings advance ongoing research on the applications of shared sequencer and time-boost~(MEV) auctions. Our key contributions are as follows.


\point{L2 Non-atomic Arbitrage and Its Decay Over Time} 
By examining the price differences between Binance and Uniswap~(v3) WETH-USDC pools on L2s, we discover over~0.5 million unexploited arbitrage opportunities on rollups. The Maximal Arbitrage Value~(MAV)~\cite{gogol2024quantifying} on Arbitrum, Base, and Optimism pools varies from~0.03\% to~0.05\% of trading volume, whereas in ZKsync Era, it is approximately~0.25\%. We show that episodes of price misalignment often last for~10--20 blocks in rollups; we further present a framework to capture only one MAV during each misalignment period, avoiding double-counting. The empirical MAV method that filters out repeated arbitrage results in~$5\times$ lower arbitrage profits compared to the Loss-Versus-Rebalancing~(LVR) metric.

\point{Cross-Rollup MEV}
We estimate the potential for cross-rollup MEV by analyzing the price disparities among WETH-USDC Uniswap~(v3) pools on major L2s. The arbitrageurs profits are influenced by the price impact (of AMM) and, thus, are the highest between the AMM pools with the highest token reserves.

\section{Related Work}
\label{sec:related_work}

There exists a substantial body of research on DeFi, particularly focused on AMM-DEXs. In particular, the systematization of knowledge (SoK) by Xu~\etal~\cite{Xu2021SoK:Protocols} and empirical investigations of Uniswap (v3) on the Ethereum blockchain~\cite{Fritsch2021ConcentratedMakers,gogol2024liquid,Heimbach2022RisksProviders,Heimbach2021BehaviorExchanges,Loesch2021ImpermanentV3} are prominent. Most of these studies evaluate the profitability of Liquidity Providers (LPs), specifically addressing impermanent loss. The analysis by Adams~\etal~\cite{adams2024dont} analyzes the costs and dynamics of swaps on Uniswap (v3) on Ethereum. Our research extends this by breaking down swap fees on rollups and assessing the influence of blobs and block time on swap dynamics. Our findings are consistent with the simulated decomposition of swap fees in rollups presented in Gogol~\etal~\cite{gogol2024crossborder}.

The Loss Versus Rebalancing metric (LVR), which compares arbitrageur profits with those of LPs, was introduced by Milionis~\etal~\cite{milionis2023automated} and empirically measured by Fritsch~\etal~\cite{fritsch2024measuring}. LVR evaluates the LP portfolio against a rebalancing portfolio that perfectly hedges the market risk of the LP position. Our investigation quantifies the absolute profit of arbitrageurs, also known as the net LVR (per block), incorporating significant adjustments. First, LVR assumes that the price discrepancies between AMMs and target exchanges, such as CEXs, disappear after each block. Our findings confirm this assumption for Uniswap~(v3) on Ethereum; however, the decay period of price discrepancies on rollups extends over multiple blocks. Our methodology adjusts the LVR to this scenario and prevents double counting of the same arbitrage opportunities. Second, LVR and the rebalancing portfolio assess the LP's loss relative to arbitrageurs, thereby containing impermanent loss. Our metrics measure the absolute profit of the arbitrageur. Practically, arbitrageurs do not replicate the rebalancing portfolio but instead capitalize on arbitrage opportunities~\cite{heimbach2024nonatomic}.

The persistence of price disparities on AMMs within ZKsync Era, a ZK rollup, was initially reported by Gogol~\etal~\cite{gogol2024quantifying}. Our paper builds on that research by providing:~i) a comprehensive analysis of rollups (Arbitrum, Base, Optimism, in addition to ZKsync Era) alongside Ethereum,~ii) enhanced precision in market price accuracy and per-block analysis, improving from a~1-minute to a~1-second resolution, and~iii) an in-depth examination of Uniswap~(v3), focusing on the most actively traded pools.

Heimbach~\etal~\cite{heimbach2024nonatomic} focus on the estimation of arbitrage executed through MEV auctions on Ethereum, specifically addressing non-atomic arbitrage opportunities between AMMs and CEXs. Our research extends this investigation by examining rollups, where MEV auctions have not yet been implemented. Similarly, we establish the minimum threshold for arbitrage opportunities and derive the arbitrage value using AMM formulas. However, our work seeks to quantify the unexploited arbitrage, which occasionally persisted for several minutes, in contrast to the arbitrage realized via MEV on Ethereum. Other studies on arbitrage~\cite{barbon2023quality,berg2022empirical,lehar2021decentralized} focus primarily on AMM-DEX on Ethereum and/or cyclic arbitrage opportunities.

Torres~\etal~\cite{torres2024rolling} is the first effort to quantify MEV in rollups. Despite the absence of MEV auctions, this research examines the heuristic for back-running arbitrage MEV, specifically focusing on swap events occurring within the same transaction in the block, within the context of cylindrical arbitrage. Our objective is to assess the potential for non-atomic arbitrage, involving CEXs, or cross-rollup price discrepancies.

Finally, Mamageishvili~\etal~\cite{mamageishvili2023buying} introduce a time boost mechanism within rollup sequencers. This mechanism facilitates back-running MEV while preventing front-running MEV. In subsequent work~\cite{mamageishvili2023shared}, a shared sequencer was proposed, enabling cross-rollup MEV auctions, such as those for arbitrage exploitation. However, these studies lack empirical quantification of arbitrage opportunities within rollups, a gap that this work seeks to address.
\section{Background}
\label{sec:backgound}


There are two main approaches~—~Layer-1 (L1) and Layer-2 (L2) scaling~—~to tackle blockchain scalability challenges and the associated increases in gas fees. L1 scaling involves creating a new blockchain with its own consensus mechanisms and a unique physical infrastructure to support the network. In contrast, L2 scaling uses a different method: it carries out computations off the main blockchain while recording their results on the underlying L1 chain. The main L2 scaling solutions include state (payment) channels, plasma, validium, and rollups. While plasma and state channels aim to move both data and computation off-chain, rollups and validium are non-custodial: they shift computation off-chain but keep compressed transaction data (rollups) or validity proofs (validiums) on the underlying L1 chain. As a result, rollups and validium require fewer trust assumptions compared to state channels, plasma chains, or L1 networks, with data security and integrity being ensured by the security of the underlying chain. 

\point{Rollup}
A rollup functions as a virtual blockchain, producing transaction and blocks and recording them on the main chain as compressed batches. However, it operates with the risk of adversarial actions such as stopping block production, generating invalid blocks, withholding data, or engaging in other malicious activities~\cite{chaliasos2024sok}. To ensure the correctness of the transaction state in these untrusted environments, rollups use one of two approaches: the optimistic approach and zero-knowledge (ZK) proofs. Figure~\ref{fig:rollup-diagram}, in~\S\ref{sec_app:backgound}, illustrates the rollup architecture, highlighting its principal components: \emph{sequencers} and \emph{verifiers}. Sequencers consolidate transactions to be rolled up onto the L1 chain, thereby achieving gas fee reductions. Verifiers, which are smart contracts on L1, validate the transaction block created by the sequencer, thus ensuring its correctness. Furthermore, ZK-rollups utilize provers responsible for generating ZK validity proofs~\cite{Chaliasos2024AnalyzingRollups}.




\point{Soft and Hard Finality}
Transaction finality refers to the point at which the state of a transaction in the blockchain ledger becomes permanent. Within the context of rollups, there are two types of transaction finality: \emph{hard finality}, which is the permanent state of a transaction on the L1 chain, and \emph{soft finality}, which is the irreversible state of a transaction on the L2 chain.~\cite{yee2022shades}

\point{Dencun Upgrade and Blobs}
On March 13, 2024, Ethereum implemented the Dencun upgrade, introducing temporary data storage called ``blobs''~\cite{2024EthereumRoadmap}. Blobs improve the efficiency of Ethereum rollups by storing data for 18 days, during which rollups validate transactions. After this period, the blobs are deleted from the blockchain. While rollups aren't required to retain blob transaction data, they usually store it off-chain to provide users with a complete transaction history.


\section{Model for Arbitrage}
\label{sec:model}

Price disparities between different trading platforms (both DEXs and CEXs) present certain arbitrage opportunities. In this section, we derive the formula for the trading volume that equals prices and maximizes the arbitrageur's profit. This value is referred to as the net LVR~\cite{milionis2023automated}, or the Maximal Arbitrage Value~(MAV)~\cite{gogol2024quantifying}. 
For more details on AMMs, we refer the reader to Appendix~\S\ref{sec_app:backgound}.

\point{Maximal Arbitrage Value (MAV)}
Consider a scenario of two tokens - X and Y - in which there is a price discrepancy between a CEX and an AMM, the CEX having a substantially higher market depth, resulting in zero market impact. Let $P_{CEX}$ and $P_{AMM}$ be the exchange prices of the token X for Y at CEX and AMM, respectively, with $g$ being the CEX fee and $f$ LP fee at AMM.

When $P_{AMM}$ < $P_{CEX}$, the arbitrageur buys $\Delta x (1-f)$ tokens X at AMM and sells them at CEX.  His profit, denominated in Y, is 

\begin{equation}
V(\Delta x) =  \Delta x (1-f) P_{CEX} (1-g) - \Delta x  P_{AMM} \rho( \Delta x )
\end{equation}

\noindent
where $\rho (\Delta x)$ is the \emph{percentage price impact} of his transaction on the AMM price.

Assuming that $P_{CEX}$ < $P_{AMM}$, the arbitrageur buys $\Delta x (1-g)$ tokens X at CEX for $\Delta x P_{CEX}$ and sells them at AMM.  His profit, denominated in Y, is 

\begin{equation}
V(\Delta x) = \Delta x (1-g) P_{AMM} ( 1 - \rho( \Delta x (1-g) )) (1-f) - \Delta x P_{CEX}
\end{equation}

\point{Arbitrage between a CPMM and CEX} 
Let us assume that the respective reserves of tokens X and Y in the AMM liquidity pool are $x$ and $y$, and $P_{AMM}$ < $P_{CEX}$.
Assuming that AMM follows CPMM, for a swap of $\Delta x $ for $\Delta y $, the percentage price impact is given by $\rho(\Delta x) = \frac{\Delta y}{y}$  \cite{gogol2024quantifying,Xu2021SoK:Protocols}. As LP fee $f \Delta x$ is not part of token reservers, then $xy=(x-\Delta x)(y+\Delta y)$ and
$1 + \frac{\Delta y}{y} = \frac{1}{1-\frac{\Delta x}{x}}$. Thus, the MAV can be expressed as: 

\begin{equation}
\label{eq:MAV_general_uniswap_v2}
V(\Delta x) = \Delta x \left[ (1-f)(1-g)P_{CEX} - \frac{P_{AMM}}{1- \frac{\Delta x}{x}} \right]
\end{equation}

\noindent
We can compute the first derivative of Eq. \eqref{eq:MAV_general_uniswap_v2} with respect to $\Delta x$ to find the max points, receiving:

\begin{equation}
\label{eq:V_max_uniswap_v2}
\Delta x_{max} = x \cdot ( 1 - \frac{\sqrt{P_{AMM}}}{\sqrt{(1-f)(1-g)P_{CEX}}} )
\end{equation}

\noindent
And, by substituting \eqref{eq:V_max_uniswap_v2} into \eqref{eq:MAV_general_uniswap_v2} the MAV is written as:
\begin{equation}
\label{eq:MAV_uniswap_v2}
 V(\Delta x_{max}) = x \cdot (1-f) \left[ \sqrt{(1-g)P_{CEX}}  -  \sqrt{\frac{P_{AMM}}{1-f}} \right] ^2
\end{equation}

\noindent
Analogously, we can find MAV when $P_{AMM} > P_{CEX}$:

\begin{equation}
\label{eq:V_max_uniswap_v2}
\Delta x_{max} = \frac{x}{1-g} \cdot ( \frac{\sqrt{(1-f)(1-g)P_{AMM}}}{\sqrt{P_{CEX}}} - 1 )
\end{equation}

\begin{equation}
 V(\Delta x_{max}) = x \cdot \left[ \sqrt{(1-f)P_{AMM}}  -  \sqrt{\frac{P_{CEX}}{1-g}} \right]^2
\end{equation}

\point{Probability of Successful Transaction} 
Let us assume that $P_{CEX} = P_{AMM} (1 + \epsilon)$, where $\epsilon > 0 $ and the probability of successful execution of the arbitrage transaction of volume $\Delta x$ is $\mathbb{P}(\Delta x, \epsilon)$. Then, the payoff of the arbitrageour is

\begin{equation}
 \hat{V}(\Delta x, \epsilon) = V(\Delta x, \epsilon) \cdot \mathbb{P}(\Delta x, \epsilon) - gas fee
\end{equation}

$$
 \hat{V}(\Delta x, \epsilon) = P_{AMM} \Delta x (1 + \epsilon - \frac{x}{x-\Delta x}) \cdot \mathbb{P}(\Delta x, \epsilon) - gas fee
$$

\point{Arbitrage between two CPMMs} 
Following analogous calculation, we can derive the formula for $x_{max}$ and MAV between two CPMMs. Assuming reserves of token X in their liquidity pools given by $x_1$ and $x_2$, and spot prices~$P_1$ and~$P_2$:

\begin{equation}
\label{eq:two_amms}
\Delta x_{max} = \frac{ \sqrt{P_1}  -  \sqrt{P_2} }{\frac{\sqrt{P_1}}{x_1}-\frac{\sqrt{P_2}}{x_2}}
\end{equation}

\section{Data Collection}
\label{sec:data}

We analyze swap data from Ethereum and its rollups with the highest trading volumes. This includes Arbitrum, Base, and Optimism (optimistic rollups), and ZKsync Era (ZK-rollup). For each blockchain, we focus on the most actively traded pool, WETH-USDC on Uniswap~(v3), during the period from December~31st,~2023 and April~30th,~2024. 

The data set is sourced from the blockchain archive nodes provided by Nansen~\cite{2024Nansen}. The number of analyzed swaps, transactions, and block ranges is detailed in Table~\ref{tab:dataset}. Using the event logs of the \emph{Swap} method, we recalculate the historical spot prices and liquidity in ticks within the USDC-WETH pools. For Uniswap~(v3), the spot price after the swap is obtained from \textit{sqrtPriceX96} in the event logs.  Market data for the ETH-USDC exchange rate on CEX are sourced from Binance APIs~\cite{Binance_API}. 

We examine 1-second Binance closing prices, noting that block production on Ethereum occurs every~12 seconds, while it is around~1--2 seconds on Base, Optimism, and ZKsync Era. On Arbitrum, block production can be as fast as~0.25 seconds. Thus, for each blockchain, except for Arbitrum, computations are performed at the block level. For Arbitrum, we analyze the last block in a second. The DEX spot price is determined by the last swap within the block.

\begin{table*}[tb]
  \centering
\resizebox{\textwidth}{!}{%
    \begin{tabular}{lrrrrrrrr}
      \toprule
      \thead{\bf Chain} & \thead{\bf Swaps} & \thead{\bf Transactions} & \thead{\bf Blocks} & \thead{\bf Block Range} \\
      \midrule
    Ethereum  & \num{761005} & \num{749818} & \num{475409} & \num{18908896}--\num{19771559} \\
    \midrule
      Arbitrum  & \num{2400000} & \num{2367361} & \num{1709619} & \num{187373628}--\num{206540031}  \\
      Arbitrum (USDC.e) & \num{2258469} & \num{2232543} & \num{1648839} & \num{165788868}--\num{206540037} \\
      Base     & \num{1687530} & \num{1636196} & \num{1145789} & \num{8638929}--\num{13866123} \\
      Optimism (USDC.e)  & \num{1186780} & \num{1136839} & \num{902789} & \num{114234215}--\num{119461410} \\
      ZKsync Era (USDC.e) & \num{46417} & \num{46379} & \num{45364} & \num{22909923}--\num{32843035}\\
      \bottomrule\\
    \end{tabular}
  }
\caption{List of analyzed WETH-USDC and WETH-USDC.e liquidity pools on Uniswap (v3) on Ethereum and its rollups, spanning the period from~31st of December~2023 to~30th of April~2024.} 
  \label{tab:dataset}
\end{table*}

\point{Native USDC and Bridged USDC.e}
USDC is a fiat-backed stablecoin issued by Circle. It takes one of two forms: native or bridged token. The native USDC is directly backed by Circle's off-chain reserves, while USDC.e is bridged, typically from Ethereum. The bridged stablecoin, USDC.e, is faster to launch on new blockchains, while the issuance of the native USDC stablecoin often follows. As a result, in most rollups, USDC and USDC.e tokens coexist for legacy purposes.

\point{Dynamics and Cost of Swapping on L2s} We break down these costs and present the characteristics of WETH-USDC swaps on rollups on Tables~\ref{tab:swap_volume} and~\ref{tab:swap_block} in~\S\ref{sec:price}, while Table~\ref{tab:swap_fees} (also in~\S\ref{sec:price}) breaks down the costs of swapping. The corresponding values of Ethereum are provided as a benchmark. As shown in Table~\ref{tab:swap_volume}, swap transactions on rollups have a lower volume compared to those on Ethereum, although they occur more frequently. On average, there are~2--3 times more swaps on rollups, but the trading volume is about five times lower. Consequently, the median swap volume on Ethereum is~\num{4239.45} USD, while it is~\num{2201.23} USD on Arbitrum,~\num{173.46} USD on Base,~\num{331.28} USD on Optimism, and only~\num{20} USD on ZKsync Era; see Appendix~\S\ref{sec:price} for details. 

\section{Empirical Results}
\label{sec:studies}

\begin{figure}[t]
\begin{subfigure}{.69\textwidth}
  \centering
    \includegraphics[width=0.8\linewidth]{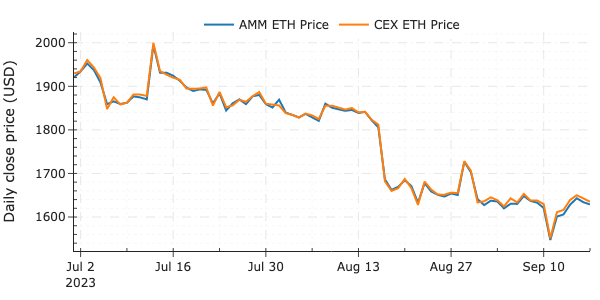}
    \caption{Time series of prices for USDC-ETH.}
    \label{fig:price-diff-evidence}
  \end{subfigure}
  \begin{subfigure}{.3\textwidth}
  \centering
    \includegraphics[width=0.8\linewidth]{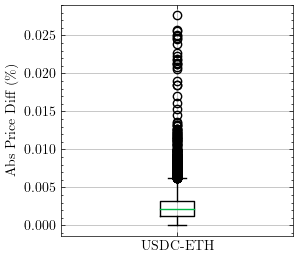}
    \caption{Distribution of deltas.}
    \label{fig:delta-price-outliers}
  \end{subfigure} \\
  \vspace{0.15cm} \\
  \centering
  \begin{subfigure}{.8\textwidth}
  \centering
    \includegraphics[width=0.8\linewidth]{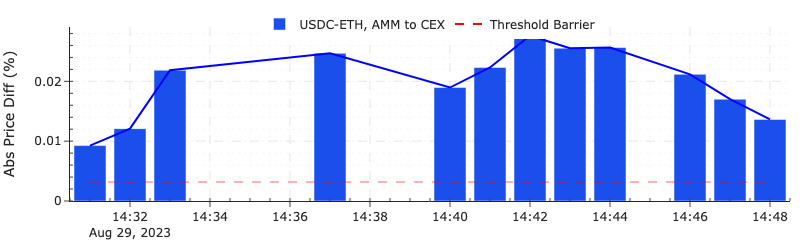}
    \caption{Price differences at the block level, with the red dashed line indicating the threshold at which prices are assumed to have re-aligned. Missing bars represent blocks where no trading activity occurred on the AMM.}
    \label{fig:explanatory-decay}
  \end{subfigure}
  \caption{These plots illustrate the price differences for USDC-ETH between the  AMM on ZKsync and Binance. The AMM spot price fluctuates around the CEX price, showing how price misalignments first widen and subsequently narrow over a period of several minutes.}
\end{figure}

When price discrepancies arise between trading platforms, arbitrageurs can take advantage of these opportunities to profit from price corrections. This activity benefits both the arbitrageurs, who gain from the price differences, and traders, who enjoy promptly adjusted and accurate exchange rates. 

This section examines the WETH-USDC price misalignment. We measure arbitrage opportunities between rollup AMMs and CEX, and across rollups. We also analyze the decay time of these discrepancies and quantify the arbitrage opportunities using the Maximal Arbitrage Value~(MAV) approach.

\subsection{Methodology}
To avoid double-counting the same arbitrage opportunities that span multiple blocks, we use the MAV methodology from~Gogol~\etal\cite{gogol2024quantifying} and adapt it for block-by-block analysis. It includes:

\point{Block intervals}
We analyze the Binance ETH-USDC closing prices at every second and compare them with the spot price on the AMM-DEX at every block. The spot price on the DEX is determined based on the last swap within the block. The block production on Ethereum takes approximately~12 seconds, followed by around~2 seconds on Base, Optimism, and ZKsync Era. On Arbitrum, blocks can be produced more quickly, approximately every~0.25 seconds. In such cases, we analyze the last block within the second. Figure \ref{fig:price-diff-evidence} shows the ETH-USDC exchange rates on Binance and ZKsync Era, highlighting the existence of price differences. 

\point{Price re-alignment threshold} 
We only consider the price deviations that surpass a set threshold. This threshold is determined from the empirical distribution of price differences (see Figure~\ref{fig:delta-price-outliers}), mainly focusing on outlier values (i.e., those beyond~1.5 times the interquartile range above the third quartile). This selected threshold ensures that prices are generally aligned but allows for adjustments to enhance flexibility.

\point{Empirical MAV}
Figure \ref{fig:explanatory-decay} illustrates the extent of price discrepancies. If a price difference continues for a prolonged duration of multiple blocks, we document only the highest MAV within each time segment of price discrepancy until re-alignment happens. This approach avoids counting the same ongoing price discrepancy multiple times.

\point{Decay time}
Once all instances of empirical MAV (i.e., the peak MAV during periods of price misalignment) are identified, we determine the duration required for prices to re-align (i.e., when the magnitude of the price difference drops below the specified threshold).


\subsection{Empirical MAV from WETH-USDC Pool}

\begin{table*}[t]
\centering
\setlength{\tabcolsep}{3.4pt}
\small
\begin{tabular}{lrrrr}
\toprule
\thead{\bf Chain} & \thead{\bf MAV} & \thead{\bf \% Volume} & \thead{\bf Net LVR} & \thead{\bf \% Volume} \\
\midrule
Ethereum   &\num{52488479.88}  & \num{0.15} & \num{126531170.12}& \num{0.36} \\
\midrule
Arbitrum   & \num{2699032.60} & \num{0.04} & \num{20686037.49}& \num{0.28} \\
Arbitrum   (USDC.e)& \num{2707008.96} & \num{0.03} & \num{19332648.22}& \num{0.24} \\
Base       & \num{1293771.96} & \num{0.04} & \num{32663704.76}& \num{1.09}  \\
Optimism   (USDC.e)& \num{690868.44}  & \num{0.05} & \num{3893401.00}& \num{0.27}\\
ZKsync Era (USDC.e)& \num{20110.31} & \num{0.25} & \num{96016.54}& \num{1.20} \\
\bottomrule\\
\end{tabular}
\caption{Assessment of arbitrage opportunities within the WETH-USDC and WETH-USDC.e liquidity pools on Uniswap~(v3) on Ethereum and its rollups from~31st of December~2023 to the~30th of April~2024. The metrics encompass the total Maximal Arbitrage Value (MAV) and net Loss-Versus-Rebalancing~(LVR), both in absolute terms and as percentages of trading volume.}
\label{tab:MAV_total}
\end{table*}

Arbitrage profit opportunities increase with price disparity, and potential profit depends on trade volume and pool reserves.
Table~\ref{tab:MAV_total} shows the total arbitrage opportunities~(MAV) during the study periods, as an absolute number and a percentage of trading volume. Although the highest price differences are on Ethereum, offering about~0.15\% of trading volume in arbitrage opportunities, these may not be fully realized due to gas fees and MEV auction costs. Additionally, specific details in the Binance order book, unavailable from 1-second data interval research, can contribute to untapped opportunities.

Non-atomic arbitrage on rollups presents significant differences from Ethereum. MAV in Arbitrum, Base, and Optimism pools ranges from~0.03\% to~0.05\% of trading volume, whereas in ZKsync, it is around~0.25\%, resulting from price misaligment occurring in multiple blocks. Empirical MAV, filtering identical arbitrage opportunities across blocks, results in lower rewards than cumulative net LVR. LVR ranges from~0.24\% to~1.2\%, about five times higher than~MAV.

\begin{table*}[t]
\centering
\setlength{\tabcolsep}{8pt}
\begin{tabular}{llrrrr}
\toprule
\thead{\bf Chain} & & \thead{\bf Points} & \thead{\bf Avg\\\bf MAV} & \thead{\bf $V_{max}$} & \thead{\bf Avg\\ \bf Decay (s)} \\
\midrule
Ethereum   & & \num{71135} & \num{737.87} & \num{499667.10} & 31 \\
\midrule
Arbitrum   & & \num{137162} & \num{19.68} & \num{56476.35} & 6.9 \\
Arbitrum (USDC.e) & & \num{164080} & \num{16.50} & \num{41370.11} & 8.8 \\
Base     &  & \num{75729} & \num{17.08} & \num{22620.17} & 420  \\
Optimism  (USDC.e)&  & \num{113428} & \num{6.09} & \num{8477.60} & 19 \\
ZKsync Era (USDC.e) &  & \num{3970} & \num{5.07} & \num{1966.76} & 370 \\
\bottomrule
\\
\end{tabular}
\caption{Arbitrage opportunities within the WETH-USDC and WETH-USDC.e liquidity pools on Uniswap (v3) on Ethereum and its rollups from Dec.~31st,~2023 to Apr.~30th,~2024. The metrics include the quantity of identified arbitrage opportunities, average Maximal Arbitrage Value~(MAV), average transaction volume yielding maximal arbitrage, and mean decay time of the price disparities.}
\label{tab:MAV_breakdown}
\end{table*}

Table~\ref{tab:MAV_breakdown} provides further insight into the opportunities analyzed. The highest quantity of arbitrage opportunities is observed in the two Arbitrum pools, which can be attributed to higher block production rates. Base and Optimism also exhibit more arbitrage opportunities compared to the Ethereum pool. In particular, the average MAV is significantly lower on rollups, approximately~5--20 USD compared to over~700 USD on Ethereum. However, the transaction volume required to achieve this arbitrage is also lower (approximately~\num{2,000}--\num{5,000}~USD on rollups and~0.5 million USD on Ethereum). Additionally, the average decay on rollups varies: for Arbitrum, it is approximately~7--9 seconds (with a~0.25-second block time), followed by~19 seconds on Optimism (2-second block time),~370 seconds on ZKsync (1.05-second block time), and~420 seconds on Base (2-second block time). In each case, the spans cover~10--20 blocks or more.

\begin{figure}[tb]
  \centering
        \includegraphics[width=0.7\linewidth]{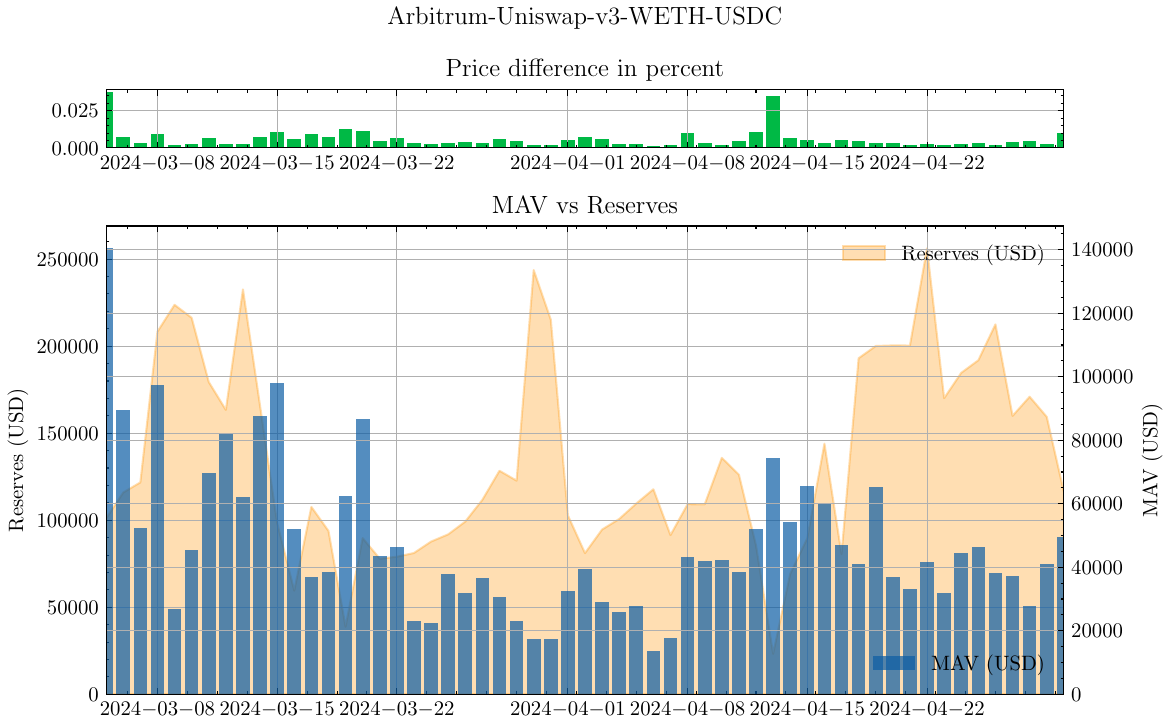}
    \caption{Empirical daily analysis of Maximal Arbitrage Value (MAV), price discrepancies, and liquidity in the current tick within the WETH-USDC liquidity pool on Uniswap (v3) deployed on the Arbitrum network.}
    \label{fig:MAV_arbitrum}
\end{figure}
\color{black}

Figure~\ref{fig:MAV_arbitrum} illustrates the changes in price over time, the MAV, and the pool's virtual reserves within the current tick for the most active pool analyzed, WETH-USDC in Arbitrum. It specifically shows the maximum daily price difference, the daily total of observed MAV, and the end-of-day liquidity in the current pool. The data reveal that the daily MAV ranges from~\$\num{2000} to~\$\num{140000}, with the lowest point occurring during low trading volumes between March and April~2024. The smaller price differences with Binance during this period result in a lower MAV. Importantly, the days with the highest daily MAV align with those with the largest price differences.

Figure~\ref{fig:MAV_Comparison} in \S\ref{sec:max_empirical} provides an analysis of MAV on Ethereum, Base, Arbitrum, and Optimism. In particular, even with the introduction of native USDC on Arbitrum and its corresponding pool on Uniswap~(v3), trading activity and associated arbitrage opportunities remain stable for both native USDC and bridged USDC tokens. A significant link between arbitrage opportunities and the size of reserves and liquidity within the current tick is observed in the ZKsync Era pool. The highest arbitrage opportunities occurred on days when liquidity within the current tick was at its maximum. In contrast, on Optimism, arbitrage opportunities are closely linked to price differences. This can be explained by the relatively small size of the pool on ZKsync Era. Despite the rise in reserves on Base, there is no corresponding increase in arbitrage profits. This indicates that the estimation of arbitrage opportunities is complex and should be evaluated based on price discrepancies, fees, and pool liquidity.


\subsection{Cross-Rollup Arbitrage}
\begin{figure}[tb]
  \centering
  \begin{subfigure}[b]{0.45\linewidth}
    \centering
    \includegraphics[width=\linewidth]{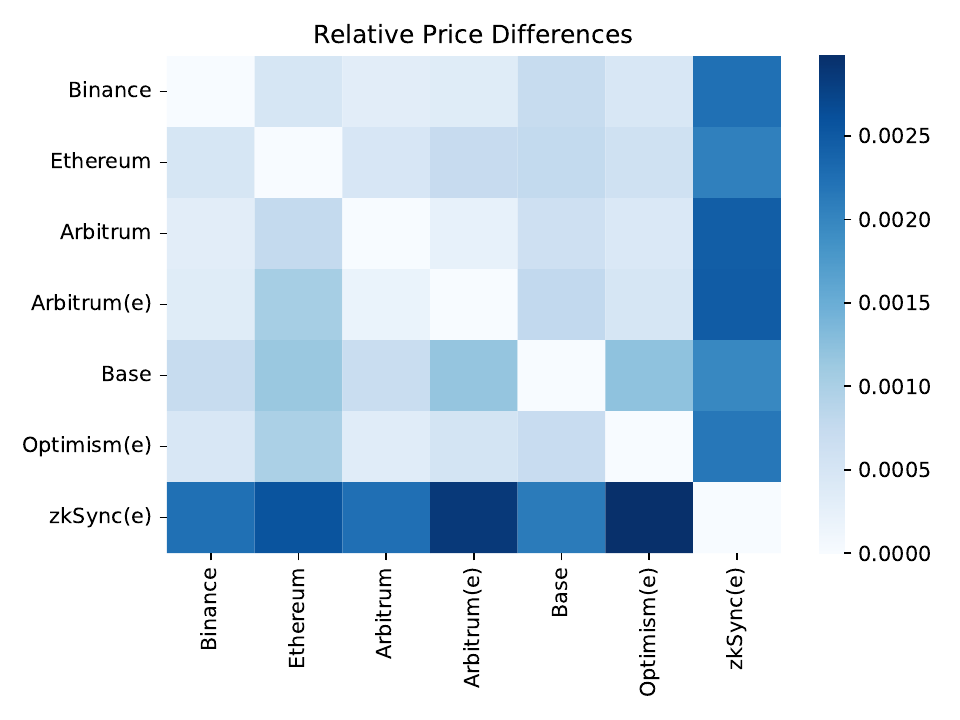}
    \caption{Mean (relative) price difference.}
    \label{fig:heatmap_price}
  \end{subfigure}
  \hfill
  \begin{subfigure}[b]{0.45\linewidth}
    \centering
    \includegraphics[width=\linewidth]{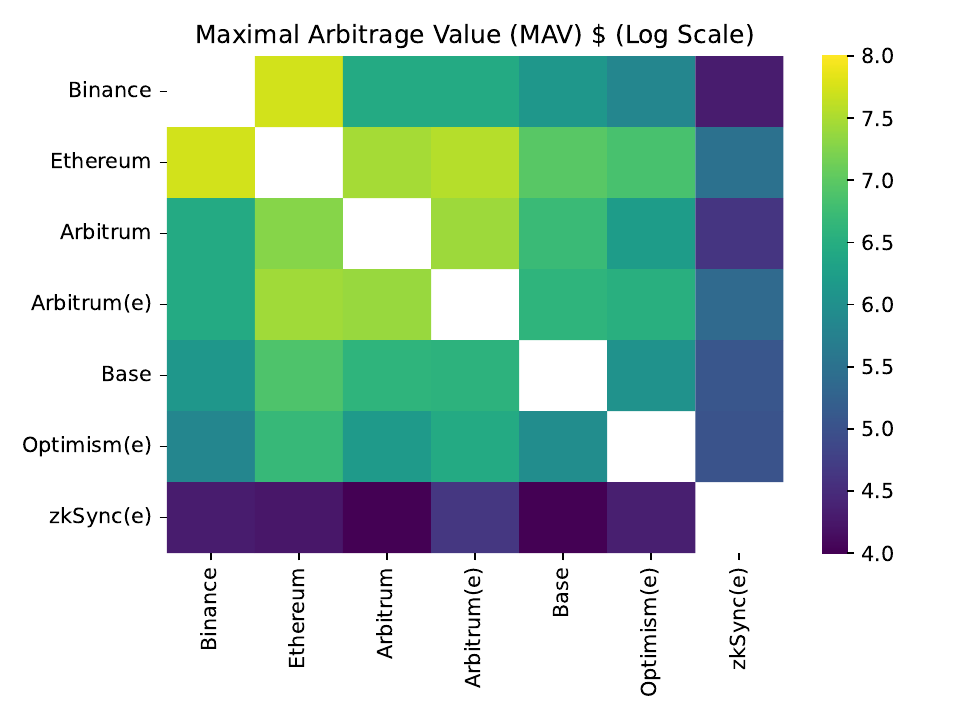}
    \caption{Maximal Arbitrage Value in USD.}
    \label{fig:heatmap_MAV}
  \end{subfigure}
  \caption{A comparative analysis of price disparities and arbitrage opportunities across rollups for the WETH-USDC and WETH-USDC.e (e) pools on Uniswap (v3). For reference, the Ethereum WETH-USDC Uniswap (v3) pool and Binance are included. }
  \label{fig:heatmap_comparison}
\end{figure}

Cross-rollup arbitrage involves carefully planned steps on at least two rollups and revenue from arbitrage must cover all fees. If an token's price is lower on one rollup than another, arbitrageurs buy low on the first rollup and sell high on the second to profit from the price difference. We show the WETH-USD price misalignment among the Uniswap~(v3) pools on different rollups in Figure~\ref{fig:heatmap_price}. For comparative analysis, we incorporate data from Binance and Uniswap (v3) on Ethereum. Figure~\ref{fig:heatmap_MAV} illustrates the MAV that could be achieved within the study period. 

%
\point{Highest discrepancy}
The most significant average price discrepancies, approximately~20 basis points~(0.0020), are observed between ZKsync and other rollups, which can be attributed to the currently limited token reserves in the Uniswap~(v3) liquidity pools on ZKsync, as illustrated in Figure~\ref{fig:MAV_Comparison} in \S\ref{sec:max_empirical}. These constrained reserves cause the highest price volatility, due to the AMM's price impact, and restrict arbitrage opportunities. Consequently, the arbitrage opportunities between ZKsync and other rollups hovered around~\num{10000} to~\num{100000} units during the period we studied.

\point{Second highest discrepancy}
The second most significant price differences with Binance are observed on Uniswap on Ethereum. This is due to Ethereum's slower block production rate of about~12 seconds, compared with~0.25 to~2 seconds on rollups. The largest arbitrage opportunities are found between Ethereum and Binance, with values reaching up to \$100 million. However, cross-rollup arbitrage opportunities exceed~\$1 million in Arbitrum, Base, and Optimism and account for about 40 basis points of trading volume, compared to 4 basis points of trading volume on Ethereum.

\point{Cross-rollup vs. DEX-to-CEX}
The cross-rollup arbitrage opportunities within the analyzed AMM pools on Arbitrum, Base, and Optimism already surpassed the value of arbitrage between these rollups and CEX. This can be attributed to the higher price volatility observed in rollup AMMs compared to~CEX.

\section{Discussion}
\label{sec:discussion}

Price disparities between trading venues reveal inefficiencies in financial systems. Untapped arbitrage on L2s may result from concerns or misunderstandings of rollup mechanics. They can be related to the centralized sequencer, rollup upgradability, bridging risk, or transaction finality. DeFi participants, particularly institutional investors, may find arbitrage opportunities less appealing compared to those available on the Ethereum network. This is attributed to the lower trading volumes and liquidity reserves observed on rollups.

\point{Layer-2 MEV}
Whereas most of the arbitrage transactions on Ethereum are executed using relayer systems such as Flashbots~\cite{Heimbach@IMC,Weintraub@IMC22}, similar mechanisms on Layer-2~(L2) solutions have not yet been developed. Furthermore, a significant proportion of MEV on Ethereum is attributed to arbitrage, with approximately~60\% being non-atomic arbitrage~\cite{dashboard@EighenPhi,dashboard@Flashbots,heimbach2024nonatomic} (between Ethereum DEX and CEX) and~40\% being cyclic (atomic) arbitrage within Ethereum. Cyclic arbitrage is already present on L2s, but it can be expected that, similar to Ethereum, the majority of arbitrage on L2 will be non-atomic, involving other rollups, Ethereum, or CEX.

\point{L2 Arbitrage Risk}
The main risks associated with L2 MEV currently stem from state (price) drift, particularly in cross-rollup arbitrage and CEX-DEX arbitrage. This drift means that price changes may occur before both legs of the transaction finalize. In contrast to cycling arbitrage, cross-rollup and DEX-CEX arbitrage are not atomic. To ensure successful transaction execution, arbitrageurs on L2s employ one of two primary strategies. Given that centralized sequencer operators manage most rollups, it is generally assumed that transactions within a block are not reorganized. Thus, the predominant strategy for arbitrageurs and MEV searchers on L2s involves partnering with the centralized sequencer to expedite transaction execution. Alternatively, another strategy entails saturating the L2 network with a high volume of potential arbitrage transactions. Leveraging the lower gas fees on L2s, this approach can be profitable. In particular, the rate of reverted transactions in rollups has increased significantly~\cite{torres2024rolling}, following the Dencun upgrade, which reduced transaction fees.

\point{Impact of a Shared Sequencer}
Engaging in cross-rollup arbitrage in the current landscape is fraught with risk. By merely submitting the transactions to the designated rollups' mempools, the arbitrageur is exposed to the risk that, by the time these transactions are sequenced, price fluctuations may occur in one or both rollups. Consequently, a subset of these transactions might fail, potentially resulting in financial losses for the arbitrageur. The quantification of this risk presents a complex problem that we leave open for future investigation.

For example, if the sequencing of transactions for both rollups $A$ and $B$ were to be managed by a single entity, the aforementioned risk could be completely mitigated. This is because the definitive state of a rollup is inherently determined by the contents of the most recent L1 block. Consequently, the proposer of the subsequent L1 block possesses exclusive authority over the progression of a rollup’s state. Specifically, the sequencer has the power to exercise discretion over which rollup transactions to include and in what sequence.

Consider the scenario where following the most recent L1 block, at height $h$, both rollups ($A$ and $B$) exhibit a state that engenders an arbitrage opportunity between their respective Uniswap pools. This particular state, and consequently the associated prices, can only be modified through the execution of transactions on the rollups. Hence, the sequencer responsible for the L1 block at height $h+1$ solely governs the subsequent progression of the rollup state for block $h+1$. Should the same sequencer be tasked with incorporating transactions for both rollups, it can ensure the execution of the arbitrage before any alteration to the rollup state occurs, for instance, by placing the arbitrage transactions at the forefront of the respective rollup's blocks.

The principal challenge within the aforementioned process lies in transferring ETH between the two rollups, which typically requires burning of ETH on rollup $A$ and minting on rollup $B$. In particular, the procedure of withdrawing from $A$ to L1 and then depositing from L1 to $B$ is likely to be excessively time consuming. Consequently, the arbitrageur is compelled to utilize a direct bridge between $A$ and $B$, capable of facilitating the cross-rollup transfer through a minimal number of transactions. The bridge, an actor relied upon (or potentially implemented)%
\footnote{The most rudimentary implementation of such a bridge involves maintaining an ETH reserve on the respective rollups.}
by the arbitrageur, holds ETH on both rollup $A$ and rollup $B$. This enables the bridge to execute the ETH transfer from $A$ to $B$ for the arbitrageur in just two transactions.

\point{Single-block finality, ZK bridging, and Flash Loans}
In the aforementioned scenario, we evaluated a trusted bridge mechanism for transferring ETH from rollup $A$ to rollup $B$. This mechanism requires the bridge to~1) maintain adequate ETH reserves on rollup $B$, and~2) authorize the bridging transaction that mints ETH on rollup $B$. This stipulation could be obviated if the rollup state were finalized instantaneously, rather than being subject to a delay as is characteristic of all prominent rollups currently in deployment. It is critical to clarify that we are addressing the finality of the rollup block relative to L1, rather than the finality of the L1 block itself. The assurance sought is that the rollup’s state can only be reverted in the event of an L1 reorganization. Now, assume that rollup $A$ operates as a ZK rollup and that a validity proof for a transaction block can be computed in a timely manner, allowing it to be included within the block itself. This facilitates the implementation of a bridge via a straightforward smart contract on rollup $B$. 

The arbitrageur can burn the ETH acquired from the swap through a transaction on rollup $A$, and the validity proof of this transaction (now incorporated in the same block) can be directly validated by the bridge contract on rollup $B$, which then mints the equivalent amount of ETH and transfers it to the arbitrageur on rollup $B$. We can extend this methodology without additional presuppositions to bridge the USDC (obtained by the arbitrageur from its subsequent swap on rollup $B$) back to rollup $A$, thereby completing the arbitrage cycle and resulting in the arbitrageur accruing more USDC on rollup $A$ than initially invested. 

The atomic execution of the entire arbitrage process presents another opportunity for the arbitrageur: securing the initial investment~(USDC) via a flash loan on rollup $A$, implying the addition of a borrowing transaction before the arbitrage and a subsequent corresponding repayment transaction. If all transactions necessary for the arbitrage (including the flash loan) are sequenced at the outset of the respective rollups’ blocks, a condition readily guaranteed by a shared sequencer, the arbitrage process is rendered entirely risk-free.

\point{Ethical MEV}
There is a debate as to whether rollup sequencers should only allow back-running MEV, as suggested by the time-boost mechanism~\cite{mamageishvili2023buying}. It is crucial to distinguish between front-running and back-running MEVs and their combined effects. Front-running MEV, especially with back-running MEV, leads to exploitative attacks like sandwich, imitation, and just-in-time liquidity attacks~\cite{zhou2021justintime}. In contrast, back-running MEV mainly supports arbitrage~\cite{qin2021quantifying} and liquidation~\cite{messias2023dissecting}, which are beneficial for DeFi by promoting price balance (arbitrage) and strengthening lending mechanisms (liquidations). 

\point{Arbitrage measures and LVR}
The LVR metric compares the profitability of arbitrageurs and LPs, whereas the MAV metric measures the arbitrage profit as net LVR per block. Furthermore, LVR assumes the price alignment within one block, so applying it to rollups results in double counting due to multi-block price discrepancies. It can be debatable whether arbitrageurs' profits should be compared with LPs' rewards. LPs earn pro-rata fees based on provided liquidity, while arbitrageurs face competition and uncertainty when executing arbitrage transformations. Thus, liquidity provision and arbitrage are usually handled by separate entities due to different infrastructure, software, and expertise.

As trading activities transition to rollups with Ethereum as a settlement layer, cross-rollup arbitrage is likely to increase, especially when rollup liquidity exceeds that of CEXs and Ethereum's DEX. Arbitrage opportunities, particularly for tokens not listed or liquid on CEXs, may arise between Ethereum and rollups.
\section{Conclusions}
\label{sec:conclusions}
This paper quantifies the potential MEV in rollups that can arise from price disparities among rollups, and between rollups and CEXs. We study WETH-USDC, the most traded cryptocurrency pair in the Ethereum ecosystem, and source prices the liquidity pools of Uniswap~(v3) on Ethereum and its L2s: Arbitrum, Base, Optimism and ZKsync Era. Binance is used as a reference CEX.

By analyzing price discrepancies after each block, we identify more than~0.5 million unexploited arbitrage opportunities. Significantly, these opportunities persist on average for a period of~10 to~20 blocks. Thus, to avoid double counting of the same opportunity, we record only the Maximal Arbitrage Value~(MAV) in each price misalignment period.  The estimated MAV on Arbitrum, Base and Optimism pools ranges from~0.03\% to~0.05\% of trading volume, and oscilates around~0.25\% in the newly-deployed Uniswap~(v3) WETH-USDC pool on ZKsync Era. The empirical MAV approach, filtering repeated price misalignments, results in five-time lower arbitrage profits compared to the Loss-Versus-Rebalancing~(LVR) metric.

We observed that lower-volume swaps are more frequent on rollups compared to Ethereum. Swaps occur~2--3 times more often on L2s, but trade volumes are approximately five times lower. Block times impact rollup transactions; Ethereum averages one swap per block, while rollups have fewer swaps per block despite higher total swaps. Specifically, swaps occur every third block on Base, fifth on Optimism, and tenth on Arbitrum. Finally, these insights contribute to ongoing discussions on rollup design and optimization, particularly in the context of MEV auctions that can be held by sequencers. 



\section*{Acknowledgements}  
The authors would like to thank Matej Pavlovic and Maria Inês Silva for their valuable insights and feedback throughout the course of this research. 

This research article is a work of scholarship and reflects the authors' own views and opinions. It does not necessarily reflect the views or opinions of any other person or organization, including the authors' employer. Readers should not rely on this article for making strategic or commercial decisions, and the authors are not responsible for any losses that may result from such use.


\bibliographystyle{splncs04}
\bibliography{references}
	
\appendix

\section{Background on Automated Market Makers}
    \label{sec_app:backgound}

\begin{figure}[t]
\centering
\includegraphics[width=0.6\textwidth]{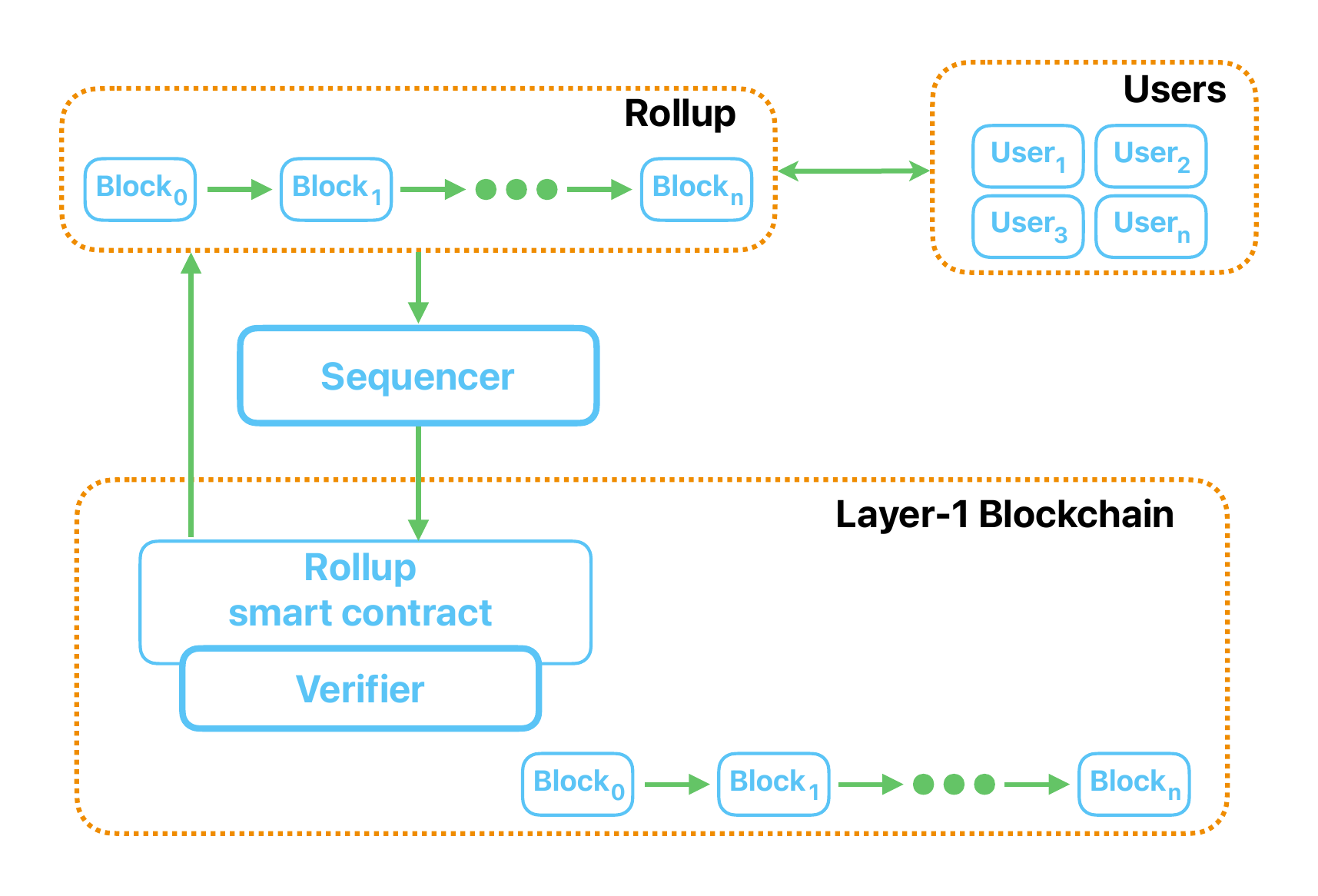} 
\caption{Architecture of a rollup.}
\label{fig:rollup-diagram}
\end{figure}

Decentralized Exchanges (DEXs) facilitate the direct exchange of tokens on the blockchain in a non-custodial and atomic way. The implementation of a Limit Order Book (LOD) on a blockchain is impractical due to storage costs and computational complexity~\cite{jensen2021AnIntroDeFi}. Consequently, DEXs employ Automated Market Makers (AMMs) to determine the exchange rate between tokens. This exchange rate is derived based on the quantity of tokens held in liquidity pools, which are supplied by Liquidity Providers (LPs)~\cite{jensen2021AnIntroDeFi}.

Constant Function Market Makers~(CFMMs) are the most common type of AMMs~\cite{Xu2021SoK:Protocols}. They use a reserve curve function $F: \mathbb{R}^N_+ \rightarrow \mathbb{R}$, which maps $N$ token reserves $x_i$ to a fixed invariant $L$. The reserve curve functions are carefully designed to improve capital efficiency within liquidity pools and to reduce the price impact for traders~\cite{Xu2021SoK:Protocols}. Each token trade on an AMM involves a trading fee, usually a fixed percentage of the swap amount, paid by the trader. These fees, or a portion of them, are then distributed proportionally among the LPs in the pool as compensation for their contribution. For simplicity, we refer to them as LP fees.

\parai{Constant Product Market Makers (CPMM).} Uniswap~(v2), one of the first and most popular AMMs, operates as a CPMM~\cite{Adams2020UniswapCore}. Its trading mechanism is governed by the formula:
\begin{equation}\label{eq:g_uniswapv2}
x_1 \cdot x_2 = L, 
\end{equation}
where $x_i$  represent the balances of the two tokens in the liquidity pool, and $L$ is a constant indicating the pool's liquidity level.

\parai{CPMM with Concentrated Liquidity (CLMM).} Uniswap~(v3) introduced concentrated liquidity, an enhancement to CPMMs, to improve the efficiency of the utilization of liquidity within the pool~\cite{Adams2021UniswapCore}. This allows LPs to specify a price range to provide liquidity. For a two-token pool ($N=2$) within the price range $[p_j,p_k]$, the trade invariant is:
\begin{equation}\label{eq:g_uniswapv3}
\left( x_1+\frac{L}{\sqrt{p_k}} \right)\left(x_2+L\cdot \sqrt{p_j}\right) = L^2,
\end{equation}
where the variables hold similar meanings as previously described. The downside of CLMMs is that LPs must continuously monitor and adjust the price range of their liquidity provision, as only trades within this range generate LP fees. Thus, LPs require liquidity provision strategies to manage LP positions and effectively minimize gas fees.

\parai{Spot Price.}
The spot exchange rate between the tokens can be determined as the slope of the $x_1-x_2$ curve, illustrated in equations~\ref{eq:g_uniswapv2} and~\ref{eq:g_uniswapv3}, with partial derivatives of the curve function $F$. 

\subsection{Cost of Swapping}
\label{sec:fees}

The total cost of executing swap transactions on an AMM includes gas costs and explicit and implicit DEX fees~\cite{Xu2021SoK:Protocols,gudgeon2020defi}. For clarity, we express this as follows.

\begin{equation}
\label{eq:cost}
\text{Total Fees} = \underset{\text{Gas Fee}}{\underbrace{\text{L1 Fee} + \text{L2 Fee}}} + \text{LP Fee} + \underset{\text{Implicit Fee}}{\underbrace{\text{Block Slippage} + \text{Price Impact}}},
\end{equation}

\point{Gas Fee} Unlike gas fees on L1 blockchains, the gas fees for rollup transactions are not predetermined as they consist of two parts: those charged by the sequencer and those charged by the underlying L1 network. The sequencer estimates the gas fee for a transaction, and this estimated amount is deducted from the originating address. Later, once the transaction, along with others, is posted on the L1 chain, the exact gas fees from L1 are determined. Depending on the specific rollup implementations, any overpayment by the transaction originator is either refunded immediately (e.g., ZKsync Era) or during the next transaction (e.g., Arbitrum).

\point{LP Fee} Also known as trading fees, LP fees are the explicit fees paid by the trader to the DEX to swap tokens. They typically range between~1 and~50 basis points (0.01\% and~0.05\%) of the swap volume, and are fully or partially redistributed to LPs.

\point{Block Slippage} This represents an implicit swap cost when trading on an AMM and can result in unexpected changes in the execution price due to the order of transactions within a block. The impact of slippage can be either positive or negative, depending on whether it benefits or disadvantages the trader in terms of the executed price. 

\point{Price Impact} Price impact is the effect of an individual trade on the market price of the underlying asset pair. It is directly related to the size of the trade and the amount of liquidity available in the AMM's pool.

\point{Arbitrage between a CLMM and CEX} Following \cite{gogol2024quantifying}, we generalize the MAV formula of CPMM - Uniswap (v3) for concentrated liquidy in Uniswap (v3). The conservation function of a Uniswap v3 pool aggregates individual LPs' conservation functions for different ticks, each dependent on the exchange rate range specified by the LP. 

Ticks occur at each price $p(i)=1.0001^i, i \in [1,2,...]$. Suppose that an LP offers liquidity ($x, y$) exclusively to users trading within a specific range of exchange rates defined by two surrounding consecutive ticks, i.e. $[P_a/\alpha, P_a\alpha]$ with $\alpha>1$ and $P_a = \frac{x}{y}$ as the current spot price. By definition, there exists some $i$ such that $[P_a/\alpha, P_a\alpha] \rightarrow [p(i), p(i+1)]$, and thus we denote for the sake of clarity ($x, y$) as ($x_i, y_i$). Within this setting, the shape of the trading function is identical to the case of liquidity provision of equivalent reserves
\begin{equation}
    x_i^{equiv} = \frac{x_i}{1-1/\sqrt{\alpha}}, \text{   and   } y_i^{equiv} = \frac{y_i}{1-1/\sqrt{\alpha}}
\end{equation}
under Uniswap v2.

We now allow the proportion of reserves to vary within the price range defined by the two ticks under consideration, and for simplicity, define $r_{i,x},r_{i,y} \geq 0$ as such varying quantities. 
Clearly, there is an upper bound to the change in reserves due to the implied movement to the next adjacent tick bounds. This translates into $0\leq r_{i,x} \leq x_i\cdot(\sqrt{\alpha}+1)$ and $0\leq r_{i,y} \leq y_i\cdot(\sqrt{\alpha}+1)^2$.
Thus, we can derive (similarly as for Uniswap v2) the percentage price impact when trading within two ticks $[p(i), p(i+1)]$ on Uniswap v3 \cite{Xu2021SoK:Protocols}. This is
\begin{equation}
    \rho_i(\Delta y) = \frac{\Delta x}{r_{i,x} + \frac{x_i}{\sqrt{\alpha}-1}},
\end{equation}
with the constraint that $\Delta y\leq r_{i,y}$, since otherwise we deplete one of the two reserves and move to the next tick.
Thus, our MAV formula becomes 
\begin{equation}
    MAV_i =
    V_{i, max} \cdot (P_a-P_c) - V_{max} \cdot P_a \cdot \rho_i( V_{i, max}),
\end{equation}
where $p(i) \leq P_c \leq p(i+1)$, and which needs to be solved enforcing the constraint $\Delta y\leq r_{i,y}$.

If the price misalignment spans multiple ticks, one then needs to iteratively compute $MAV_i$ and sum the related profits until realignment, i.e.
\begin{equation}
    MAV_{Uniswap\_v3} = \sum_{i:P_a\in [p(i), p(i+1)]}^{i:P_c\in [p(i), p(i+1)]} MAV_i.
\end{equation}

\begin{figure}[t]
  \centering
  \resizebox{1.1\linewidth}{!}{ 
    \begin{tabular}{cc}
      \begin{subfigure}[b]{0.49\linewidth}
        \centering
        \includegraphics[width=\linewidth]{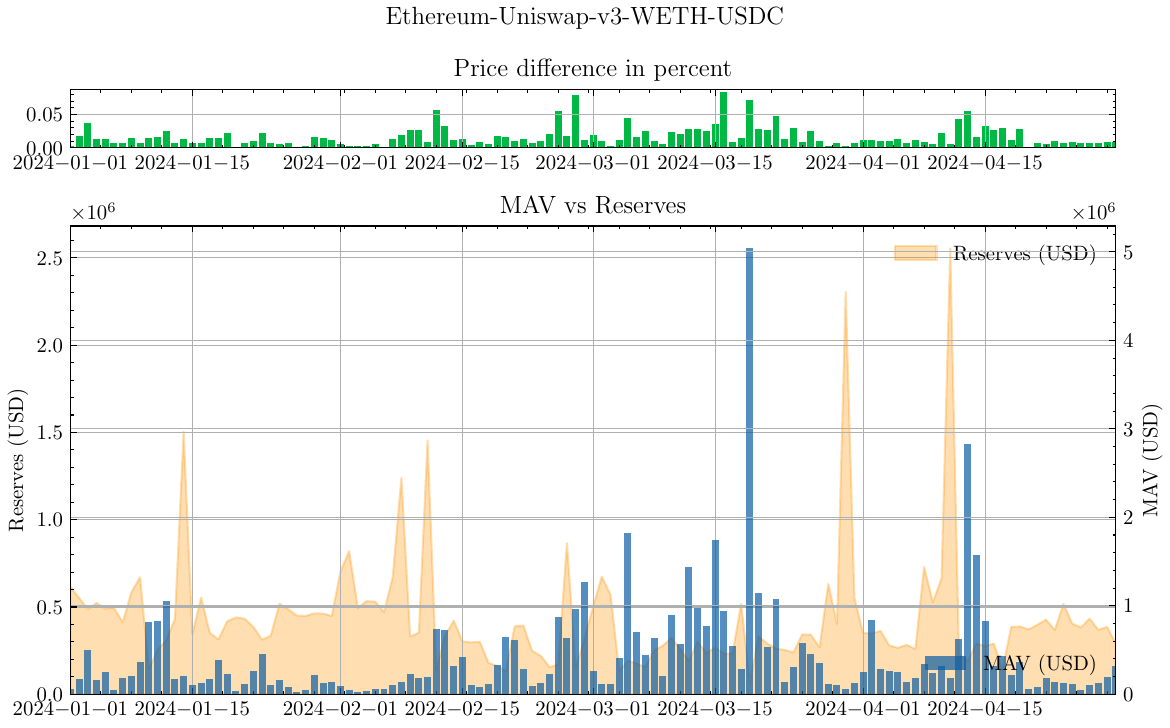}
        \caption{\small Ethereum}
        \label{fig:MAV_Ethereum}
      \end{subfigure} &
      \begin{subfigure}[b]{0.49\linewidth}
        \centering
        \includegraphics[width=\linewidth]{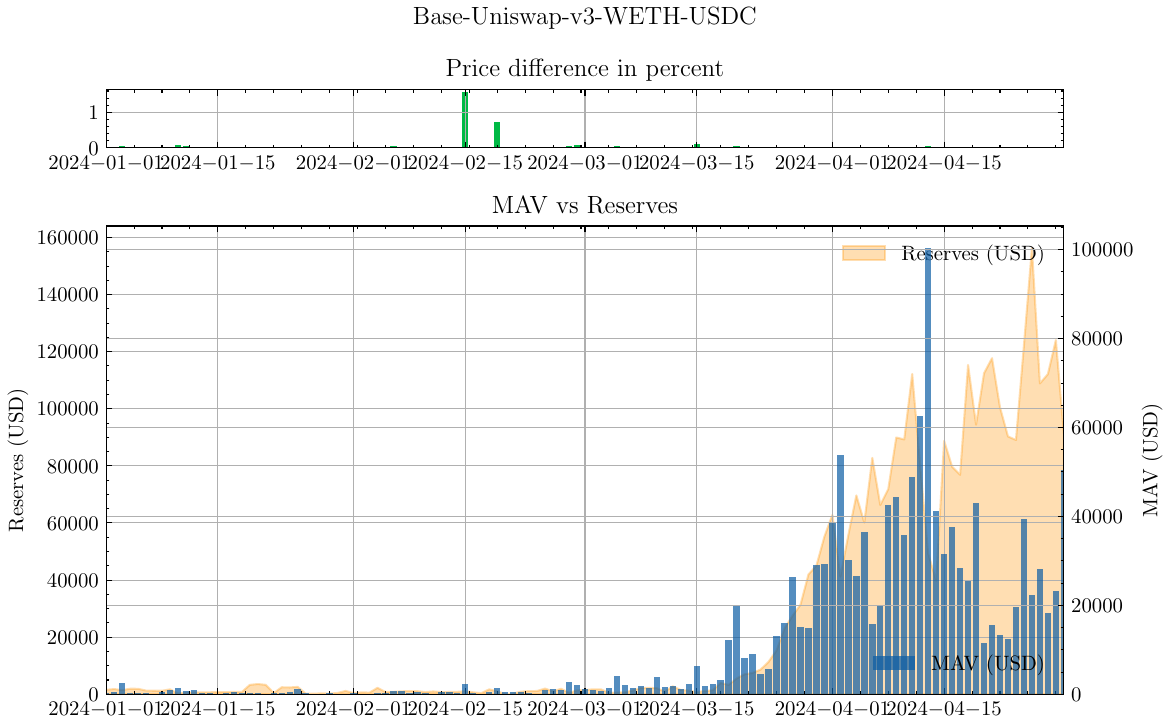}
        \caption{\small Base}
        \label{fig:MAV_Base}
      \end{subfigure} \\
      \begin{subfigure}[b]{0.49\linewidth}
        \centering
        \includegraphics[width=\linewidth]{figures/arbitrum-Uniswap-v3-WETH-USDC_mav.pdf}
        \caption{\small Arbitrum}
        \label{fig:MAV_Arbitrum1}
      \end{subfigure} &
      \begin{subfigure}[b]{0.49\linewidth}
        \centering
        \includegraphics[width=\linewidth]{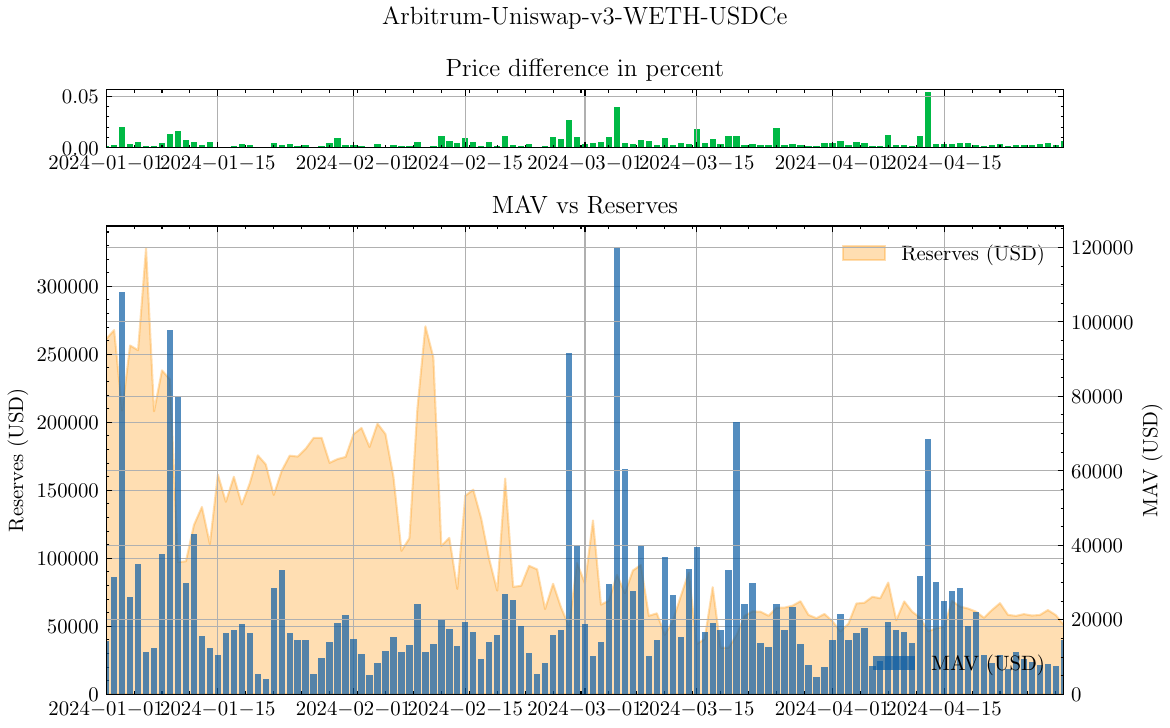}
        \caption{\small Arbitrum (USDC.e)}
        \label{fig:MAV_Arbitrum2}
      \end{subfigure} \\
      \begin{subfigure}[b]{0.49\linewidth}
        \centering
        \includegraphics[width=\linewidth]{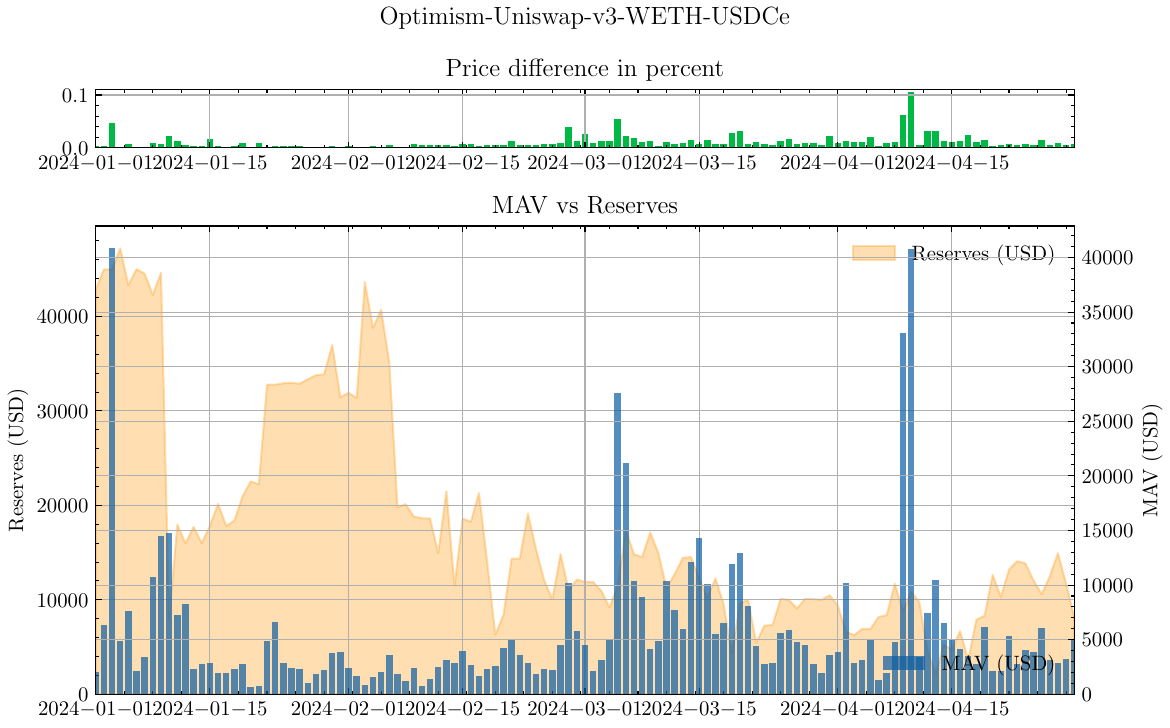}
        \caption{\small Optimism (USDC.e)}
        \label{fig:MAV_Optimism}
      \end{subfigure} &
      \begin{subfigure}[b]{0.49\linewidth}
        \centering
        \includegraphics[width=\linewidth]{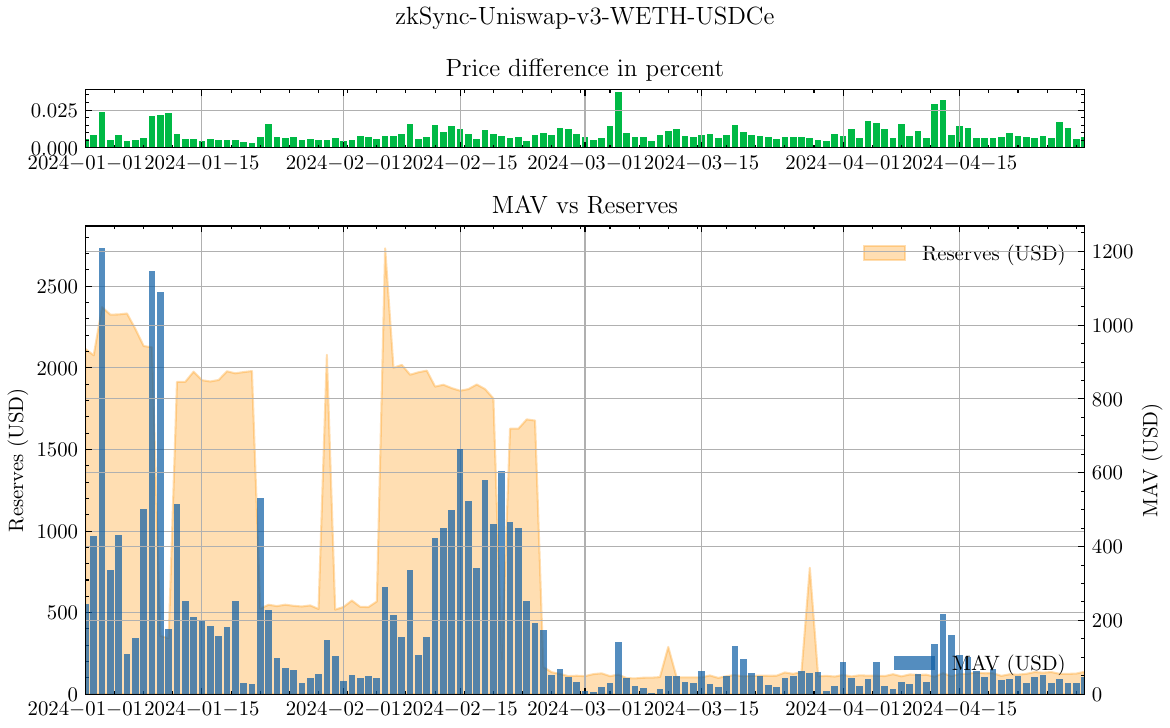}
        \caption{\small ZKsync Era (USDC.e)}
        \label{fig:MAV_zkSync}
      \end{subfigure}
    \end{tabular}
  }
  \caption{Empirical Maximal Extractable Value (MAV), price misalignment with Binance, and liquidity within the spot price tick of the WETH-USDC pool on Uniswap~(v3) across Ethereum and its rollups from December~31st,~2023 to April~30th,~2024.}
  \label{fig:MAV_Comparison}
\end{figure}

\section{Dynamics and Cost of Swapping on L2s}
\label{sec:price}

\begin{table*}[t]
\centering
\setlength{\tabcolsep}{7pt}
\begin{tabular}{lrrrrr}
\toprule
\thead{\bf Chain} & \thead{\bf Swaps} & \thead{\bf Total\\\bf Volume\\\bf mn} & \thead{\bf Avg\\ \bf Swap\\\bf Volume} & \thead{\bf Std\\\bf Swap\\\bf Volume} & \thead{\bf Median\\\bf Swap\\\bf Volume} \\
\midrule
Ethereum & \num{761005} & \num{35175} & \num{46222.26} & \num{142890.62}& \num{4239.45} \\
\midrule
Arbitrum  & \num{2400000} & \num{7266}  & \num{3027.66} & \num{6738.06}& \num{1125.01} \\
Arbitrum (USDC.e) & \num{2258469} & \num{8159} & \num{3612.86} & \num{5976.01}& \num{2201.23} \\
Base   & \num{1687530} & \num{2996} & \num{1775.55} & \num{5611.25}& \num{173.46} \\
Optimism (USDc.e) & \num{1186780} & \num{1457} & \num{1228.30} & \num{2842.73}& \num{331.28} \\
ZKsync (USDC.e) & \num{46417} & \num{8.01} & \num{172.67} & \num{426.52}& \num{20.99} \\
\bottomrule\\
\end{tabular}
\caption{An analysis of the absolute swap volumes within the WETH-USDC and WETH-USDC.e liquidity pools on Uniswap~(v3) across the Ethereum network and its rollups, spanning the period from~31st of December~2023 to~30th of April~2024.}
\label{tab:swap_volume}
\end{table*}

Tables~\ref{tab:swap_volume} and~\ref{tab:swap_block} present the characteristics of WETH-USDC swaps on rollups, while Table ~\ref{tab:swap_fees} breaks down the costs of swapping. The corresponding values of Ethereum are provided as a benchmark. As shown in Table~\ref{tab:swap_volume}, swap transactions on rollups have a lower volume compared to those on Ethereum, although they occur more frequently. On average, there are~2--3 times more swaps on rollups, but the trading volume is about five times lower. Consequently, the median swap volume on Ethereum is~\num{4239.45} USD, while it is~\num{2201.23} USD on Arbitrum,~\num{173.46} USD on Base,~\num{331.28} USD on Optimism, and only~\num{20} USD on ZKsync Era. 

Table~\ref{tab:swap_block} explains the effect of block times on rollup swap transactions. Despite having a higher number of swaps, there are fewer swaps per block on rollups because of faster block production. In Ethereum, there are, on average,~\num{0.88} swaps in every block, with every block containing swap transactions having~1.6 swap events. In contrast, every third block on Base, fifth block on Optimism, and tenth block on Arbitrum contain a swap. Furthermore, there are fewer swap events per block with swap transactions (approximately~1--1.47) on rollups. 

On average, there are slightly more swap events per transaction~(1.01) on both rollups and Ethereum, indicating that some swaps are performed in batches. This could imply the exploitation of MEV by certain wallets or decentralized exchange (DEX) aggregators initiating swaps on behalf of users.

Table~\ref{tab:swap_fees} shows the costs of swapping on L2s before and after the Dencun upgrade. Although the otal costs and average gas fees decreased, the percentage costs of swaps often remained unchanged or increased. This is due to the smaller swap sizes on rollups post-upgrade.

\begin{table*}[t]
  \centering
  \setlength{\tabcolsep}{4pt}
  \makebox[\textwidth][c]{
    \begin{tabular}{lrrrr}
      \toprule
      \thead{\bf Chain} &  \thead{\bf Block\\\bf Time (s)\\} & \thead{\bf Swaps\\ \bf per Block} & \thead{\bf Swaps per\\ \bf Transaction} & \thead{\bf Swaps Per Block\\\bf with Transaction} \\
      \midrule
     Ethereum &  12.12 & \num{0.88} & \num{1.01}  & \num{1.60} \\
    \midrule
      Arbitrum &  0.25 & \num{0.13} & \num{1.01}  & \num{1.40}  \\
      Arbitrum (USDC.e) &  0.25 & \num{0.06} & \num{1.01}  & \num{1.37}  \\
      Base  &  2.00 & \num{0.32} & \num{1.03}  & \num{1.47}  \\
      Optimism (USDC.e)  &  2.00 & \num{0.23} & \num{1.04}  & \num{1.31}  \\
      ZKsync Era (USDC.e) & 1.05 & \num{0.00} & \num{1.00}  & \num{1.02}  \\
      \bottomrule\\
    \end{tabular}
  }
 \caption{An analysis of the swaps per transactions and blocks within the WETH-USDC and WETH-USDC.e liquidity pools on Uniswap (v3) on the Ethereum network and its rollups, spanning the period from the~31st of December~2023 to the~30th of April~2024.} 
  \label{tab:swap_block}
\end{table*}

\begin{table*}[t]
\centering\footnotesize
\setlength{\tabcolsep}{2pt}
\begin{tabular}{lcccrrr}
\toprule
\setlength{\tabcolsep}{6pt}
\thead{\bf Chain}  & \thead{\bf Total Fee} & \thead{\bf Gas Fee} & \thead{\bf LP Fee} & \thead{\bf Slippage and\\ \bf Price Impact} \\
\midrule
\textit{Whole Period:} &&           &         &        &        &              \\
\midrule
Ethereum  & \num{97.20} (21bps) & \num{37.13} (8bps) & \num{23.11} (5bps) & \num{65.33} (14bps) \\
\midrule
Arbitrum & \num{2.46} (8bps) & \num{0.65} (2bps) & \num{1.51} (5bps) & \num{1.82} (6bps) &  \\
Arbitrum (USDC.e) & \num{5.55} (15bps) & \num{3.41} (9bps) & \num{1.81} (5bps) & \num{2.18} (6bps) \\
Base      & \num{1.76} (10bps) & \num{0.52} (3bps) & \num{0.89} (5bps) & \num{1.24} (7bps)\\
Optimism (USDC.e)  & \num{1.33} (11bps) & \num{0.31} (2bps) & \num{0.61} (5bps) & \num{1.10} (9bps) \\
ZKsync (USDC.e)    & \num{0.87} (50bps) & \num{0.25} (15bps) & \num{0.52} (30bps) & \num{0.65} (38bps) \\
\midrule
\textit{Pre-Blobs:}  &&           &         &        &                &              \\
\midrule
Ethereum  & \num{92.96} (22bps) & \num{39.30} (9bps) & \num{20.94} (5bps) & \num{59.43} (14bps)  \\
\midrule
Arbitrum   & \num{9.68} (10bps) & \num{4.16} (4bps) & \num{4.73} (5bps) & \num{5.76} (6bps) \\
Arbitrum (USDC.e)  & \num{6.28} (11bps) & \num{3.08} (6bps) & \num{2.77} (5bps) & \num{3.30} (6bps) \\
Base     & \num{0.43} (22bps) & \num{0.11} (6bps) & \num{0.10} (5bps) & \num{0.36} (18bps) \\
Optimism (USDC.e)   & \num{1.55} (10bps) & \num{0.36} (2bps) & \num{0.78} (5bps) & \num{1.27} (8bps) \\
ZKsync (USDC.e)  & \num{1.37} (49bps) & \num{0.37} (13bps) & \num{0.83} (30bps) & \num{1.03} (37bps) \\
\midrule
\textit{Post-Blobs:} &&           &         &        &                &              \\
\midrule
Ethereum  & \num{102.26} (20bps) & \num{34.56} (7bps) & \num{25.69} (5bps) & \num{72.35} (14bps) \\
\midrule
Arbitrum   & \num{2.00} (8bps) & \num{0.42} (2bps) & \num{1.31} (5bps) & \num{1.57} (6bps)  \\
Arbitrum (USDC.e) & \num{5.05} (22bps) & \num{3.64} (16bps) & \num{1.15} (5bps) & \num{1.41} (6bps) \\
Base      & \num{1.99} (10bps) & \num{0.59} (3bps) & \num{1.02} (5bps) & \num{1.40} (7bps) \\
Optimism (USDC.e)  & \num{1.17} (12bps) & \num{0.26} (3bps) & \num{0.49} (5bps) & \num{0.97} (10bps) \\
ZKsync (USDC.e)  & \num{0.32} (56bps) & \num{0.12} (21bps) & \num{0.17} (30bps) & \num{0.24} (40bps)  \\
\bottomrule
\end{tabular}
\caption{Breakdown of swap fees at the WETH-USDC and WETH-USDC.e pools at Uniswap (v3) at Ethereum and its rollups during the period 12.31-30.04, 13.03-30.04.24 (pre-blobs) and 14.3-30.04.24 (post-blobs).}
\label{tab:swap_fees}
\end{table*}


\section{Empirical Maximal Extractable Value (MAV)}
    \label{sec:max_empirical}

Figure~\ref{fig:MAV_Comparison} shows an analysis of MAV across Ethereum, Base, Arbitrum, and Optimism. Notably, despite the introduction of native USDC on Arbitrum and its corresponding pool on Uniswap~(v3), trading activity and arbitrage opportunities have remained steady for both native and bridged USDC tokens.
In the ZKsync Era pool, a strong relationship is observed between arbitrage opportunities and the size of reserves and liquidity within the current tick. The most significant arbitrage opportunities occurred when liquidity within the current tick reached its peak. Conversely, on Optimism, arbitrage opportunities are more closely tied to price discrepancies, likely due to the relatively small pool size on ZKsync Era.
Interestingly, while reserves have increased on Base, this hasn't led to a proportional rise in arbitrage profits. This suggests that arbitrage potential is complex and must be evaluated not only by price differences but also by considering fees and pool liquidity.

\end{document}